\documentclass[aip,apl,twocolumn,groupedaddress,reprint]{revtex4-1}

\usepackage{amssymb}
\usepackage{amsmath}
\usepackage{graphicx}
\usepackage{natbib}
\usepackage{epstopdf}

\newcommand{\unit}[1]{\mathrm{\ #1}}

\begin{document}
\textheight = 63\baselineskip
\title{Highly-tunable formation of nitrogen-vacancy centers via ion implantation}

\author{S. Sangtawesin}
\author{T. O. Brundage}
\author{Z. J. Atkins}
\author{J. R. Petta}
\affiliation{Department of Physics, Princeton University, Princeton, NJ 08544, USA}
\date{\today}

\begin{abstract}
We demonstrate highly-tunable formation of nitrogen-vacancy (NV) centers using 20 keV $^{15}$N$^+$ ion implantation through arrays of high-resolution apertures fabricated with electron beam lithography. By varying the aperture diameters from $80$ to $240 \unit{nm}$, as well as the average ion fluences from $5\times 10^{10}$ to $2\times 10^{11} \unit{ions/cm^2}$, we can control the number of ions per aperture. We analyze the photoluminescence on multiple sites with different implantation parameters and obtain ion-to-NV conversion yields of $6-7$ \%, consistent across all ion fluences. The implanted NV centers have spin dephasing times $T_2^*$ $\sim$ 3 $\mu$s, comparable to naturally occurring NV centers in high purity diamond with natural abundance $^{13}$C. With this technique, we can deterministically control the population distribution of NV centers in each aperture, allowing for the study of single or coupled NV centers and their integration into photonic structures.
\end{abstract}


\maketitle

Advances in quantum information processing (QIP) require the use of multiple qubits that are stable and easily addressable. The NV center in diamond stands out as a candidate for this application due to its spin-dependent fluorescence and long coherence time at room temperature.\cite{Balasubramanian_NMat_8_2009,Waldherr_Nature_2014,Taminiau_NatNano_2014,Dolde_NatPhys_2013} However, the feasibility of integrating naturally occurring NV centers into a large-scale QIP architecture is limited by their random locations in the diamond lattice. The technique of nitrogen ion implantation can overcome this obstacle by offering precise control of NV center locations, while maintaining the quality of the NV centers created.\cite{Dolde_NatPhys_2013,Meijer_APL_2005,Spinicelli_NJP_2011,Osterkamp_APL_2013,Yamamoto_PhysRevB.88.201201,Chu_NanoLett_2014,Pezzagna_PSSA201100455,Lesik_PSSA_2013}

In order to achieve high accuracy placement of NV centers, techniques such as implantation through a scanning force microscope tip, focused-ion beam, and apertures in implantation masks have been developed.\cite{Weis_2008,Pezzagna_PSSA201100455,Lesik_PSSA_2013} In terms of ease of fabrication and scalability, one of the most versatile methods is implantation through lithographically defined apertures.\cite{Toyli_NanoLett_2010,Spinicelli_NJP_2011}

In this Letter, we study the efficacy of this method by demonstrating highly-controllable NV implantations with different ion fluences across a wide range of aperture diameters. Within each ion fluence, aperture diameters vary from 80 to 240 nm. We characterize the implanted NV centers using photoluminescence (PL) data and autocorrelation measurements $g^{(2)}(\tau)$.\cite{Scully_QuantumOptics} Together, these data allow us to determine the statistics of NV center formation. We observe a linear relationship between the mean number of NV centers per aperture and the aperture area, from which we extract implantation yields of $6-7\%$. These yields are consistent across all ion fluences and with previously reported values.\cite{Spinicelli_NJP_2011,Toyli_NanoLett_2010,Pezzagna_NJP_2010} Finally, we measure spin dephasing times $T_2^* \sim 3 \unit{\mu s}$, a value comparable to that of naturally occurring NV centers, thus demonstrating the capability to maintain high quality NV centers and fine-tune the NV population distribution at well-defined locations.\cite{Childress_Science_2006,vanderSar_Nature_2012}

\begin{figure} [tr!]
\begin{center}
		\includegraphics[width=\columnwidth]{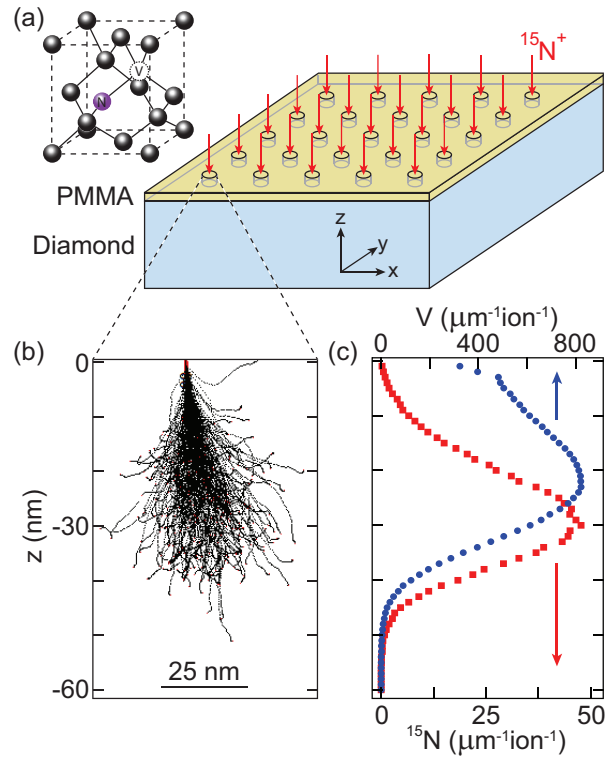}
\caption{ (Color online) (a) Implantation method: A 230 nm thick layer of PMMA serves as an implantation mask. Arrays of apertures are patterned using electron beam lithography. (b) SRIM simulation of $^{15}\mathrm{N}^+$ ion implantation with 20 keV energy. (c) Distribution of $^{15}$N atoms and vacancies, $V$, per micron of depth as a function of implantation depth, $z$, integrated over the $xy$--plane.}
\end{center}	
\end{figure}

\begin{figure*} [tr!]
\begin{center}
		\includegraphics[width=2\columnwidth]{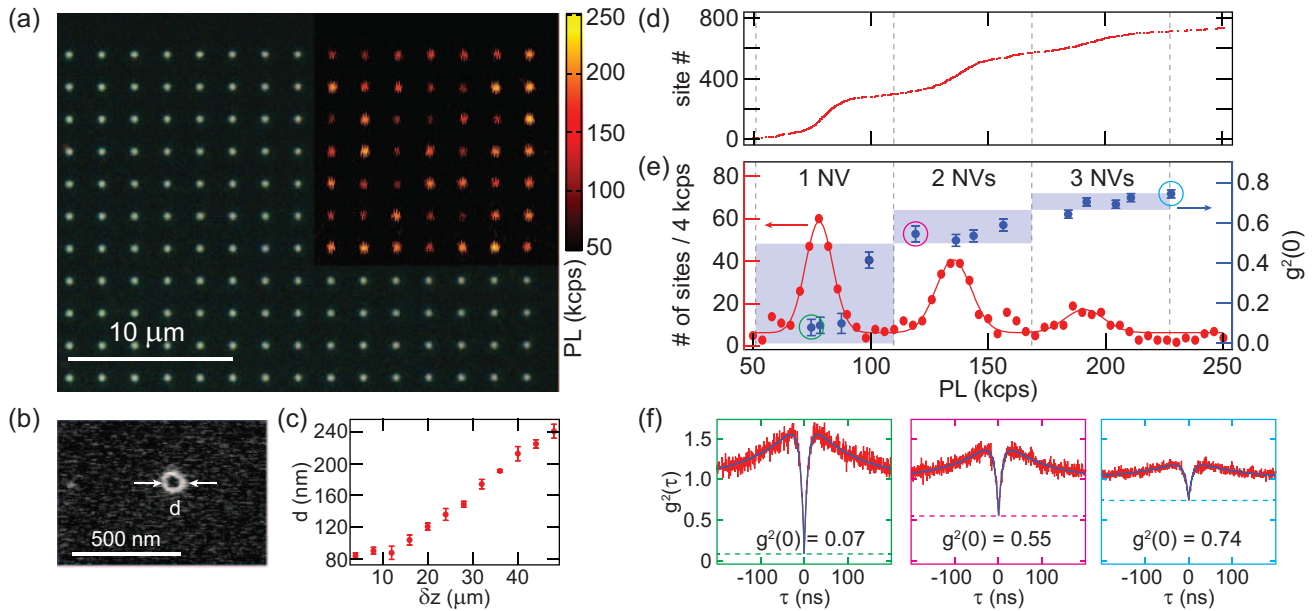}
\caption{ (Color online) (a) Optical dark field image of the PMMA implantation mask. Overlay: A confocal microscope image shows PL from implanted NV centers (with the photon count rate in units of kilocounts per second, kcps). (b) Scanning electron micrograph of an aperture with diameter $d = 82 \unit{nm}$. (c) Aperture diameter $d$ as a function of the electron beam focus offset $\delta z$. (d) PL data from 750 implantation sites showing distinct count rates associated with different numbers of NVs. (e) Red data points and curves show histograms of PL from (d) and Gaussian fits to the data, respectively. Expected ranges of the autocorrelation minima $g^{(2)}(0)$ for one, two, and three single-photon emitters are defined by the blue shaded regions. Blue data points correspond to measured values of $g^{(2)}(0)$, indicating that we have one, two, and three NVs in those sites. (f) Autocorrelation function $g^{(2)}(\tau)$ from the one-, two-, and three-NV sites indicated by the circles in (e).}
\end{center}	
\end{figure*}

We begin by selecting an electronic grade diamond ($\mathrm{N} < 5 \unit{ppb}$, natural abundance $^{13}$C, Element Six) with (100) orientation and low background PL. Typically, no NV centers are observable within our confocal microscope's $60\times 60\unit{\mu m^2}$  field of view before implantation. As illustrated in Fig.\ 1(a), we create an implantation mask on the diamond surface by spin coating the diamond with a $t$ = $230\unit{nm}$ thick layer of PMMA electron beam resist. Arrays of apertures are patterned on this PMMA mask in $100 \times 100 \unit{\mu m^2}$ grids with $2 \unit{\mu m}$ pitch using 125~kV electron beam lithography. Following the lithography, we implant the sample with $20\unit{keV}$ $^{15}\mathrm{N}^+$ ions at a $7^\circ$ tilt to prevent ion channeling.\cite{Spinicelli_NJP_2011,Naydenov_APL_2010}

We simulate the implantation process using Stopping and Range of Ions in Matter (SRIM).\cite{Ziegler20101818,Toyli_NanoLett_2010} We use a diamond substrate density of $3.52 \unit{g/cm^3}$ and a displacement energy of $37.5\unit{eV}$.\cite{Toyli_NanoLett_2010} Figure 1(b) shows the trajectories resulting from the implantation of 1,000 ions, while Fig.\ 1(c) provides the statistical distributions of the implanted $^{15}\mathrm{N}$ and the vacancies that are created due to implantation. The average depth of implanted $^{15}\mathrm{N}$ is $\sim 30 \unit{nm}$ for our implantation energy of $20\unit{keV}$. A simulation performed with a PMMA target shows that the ions are stopped with $>99.99\%$ probability within the PMMA layer.

In order to obtain reliable results, the same diamond sample is used for all implantations. Three separate implantations are performed by exposing the masked substrate to a $^{15}$N$^+$ ion beam with average fluences of $2\times 10^{11}$, $1\times 10^{11}$, and $5\times 10^{10} \unit{ions/cm^2}$ in each exposure, respectively. The average fluence in each exposure is controlled by the beam current and the exposure time. After the last implantation, the sample is cleaned in a boiling mixture of 1:1:1 nitric, perchloric, and sulfuric acid for 30 minutes. The sample is then annealed at $850 \unit{^\circ C}$ in vacuum for 2 hours to mobilize the vacancies, repair lattice damage, and allow the vacancies to be captured by the substitutional nitrogen atoms, forming NV centers.\cite{Mainwood_PhysRevB.49.7934,Twitchen_PhysRevB.59.12900,footnote_amorphous,Uzan-Saguy_APL_1995} A second acid cleaning step is performed for 4 hours after annealing to remove graphitic carbon and to oxygen terminate the surface.\cite{Naydenov_APL_2010,Antonov_APL_2014,Chu_NanoLett_2014}

To control the number of ions implanted through different apertures during one ion exposure, we vary the aperture diameter by shifting the sample out of the focal plane of the electron beam during the lithography process. With the sub-micron positioning accuracy and the large depth of focus ($\sim 10 \unit{\mu m}$) of the electron beam, this technique results in reproducible and highly-tunable aperture diameters. Figure 2(a) shows an optical dark field image of the PMMA implantation mask with $\sim 200\unit{nm}$ diameter apertures separated by $2\unit{\mu m}$. We image a sample subset of apertures using a scanning electron microscope (SEM) to determine the aperture diameter, $d$, as shown in Fig.\ 2(b).\cite{footnote_SEM} Figure 2(c) shows $d$ measured as a function of the electron beam focus offset, $\delta z$, demonstrating the ability to tune the aperture diameters from 80 to 240 nm.

The implanted sample is characterized using a scanning confocal microscope.\cite{Sangtawesin_arxiv_2014} A 532 nm excitation laser is focused onto a diffraction-limited spot on the sample with a high numerical aperture objective lens and the resulting PL is measured with an avalanche photodiode operating in the single-photon counting regime. The PL of the NV centers, represented by the photon count rate, is dependent on the polarization of the excitation photons. For a (100)-oriented diamond surface, there is a two-fold degeneracy of the four NV axes when projected onto the (100) face.\cite{Epstein_NatPhys_1_2005} To produce consistent PL levels across all sites regardless of NV center orientation, we use a circularly polarized $532 \unit{nm}$ laser excitation, along with a laser power high enough to saturate the cycling transitions for multiple NVs ($\sim 2 \unit{mW}$ at the objective). As shown in Fig.\ 2(a), measurements of the PL show emission from the implanted array of NV centers. The PL data from several implantation sites are carefully analyzed for each ion exposure and aperture diameter to determine the implantation efficiency.

Figure 2(d) shows raw PL data from a sample of 750 sites implanted with average ion fluences of $2\times 10^{11}$ and $1\times 10^{10} \unit{ions/cm^2}$. We estimate the average background PL level to be $20-30 \unit{kcps}$. Since NV centers are single-photon emitters, we expect the photon emission rate from each site to be proportional to the number of NVs within that site. The dense population around 80, 135, and 190 kcps, as indicated by the histogram in Fig.\ 2(e), is suggestive of distinct PL levels for one, two, and three NV centers, respectively.

To confirm that the PL levels from the histogram do indeed correspond to discrete numbers of NV centers, the photon autocorrelation function $g^{(2)}(\tau)$ is measured on a subset of implantation sites. Blue data points in Fig.\ 2(e) show that the autocorrelation minima $g^{(2)}(0)$ are in excellent agreement with their expected values, which are indicated by the shaded blue regions. The expected values of $g^{(2)}(0)$ are given by
\begin{equation}
1-\frac{1}{n} < g^{(2)}(0) < 1-\frac{1}{n+1} \label{eq:g2}
\end{equation}
for an $n$-photon source. Photon bunching effects from the high excitation power are also observed for larger delay times $\tau$, as shown in Fig.\ 2(f), where $g^{(2)}(\tau) > 1$.\cite{Scully_QuantumOptics,Hausmann_NJP_2011}

Using the now established scaling of the PL to the number of NV centers, we obtain an average number of NV centers per site, $\bar{n}_\mathrm{NV}$, by analyzing PL data from $100-400$ implantation sites for each implantation parameter (ion exposure and focus offset). Figure 3(a) shows the resulting $\bar{n}_\mathrm{NV}$, obtained from the three ion exposures, as a function of the effective aperture area, $A$, given by
\begin{equation}
	A = \frac{d^2}{2}\left( \cos^{-1} (\beta \tan\theta) - \beta \tan\theta \sqrt{1-\beta^2\tan^2\theta} \right). \label{eq:shadow}
\end{equation}
This takes into account shadowing from the PMMA mask due to the $\theta = 7^\circ$ implantation angle and the aspect ratio of the aperture $\beta = t/d$. The effective aperture area is $\sim 15-50\%$ smaller than the physical area of the aperture for our range of aperture diameters.

\begin{figure} [tr!]
\begin{center}
		\includegraphics[width=\columnwidth]{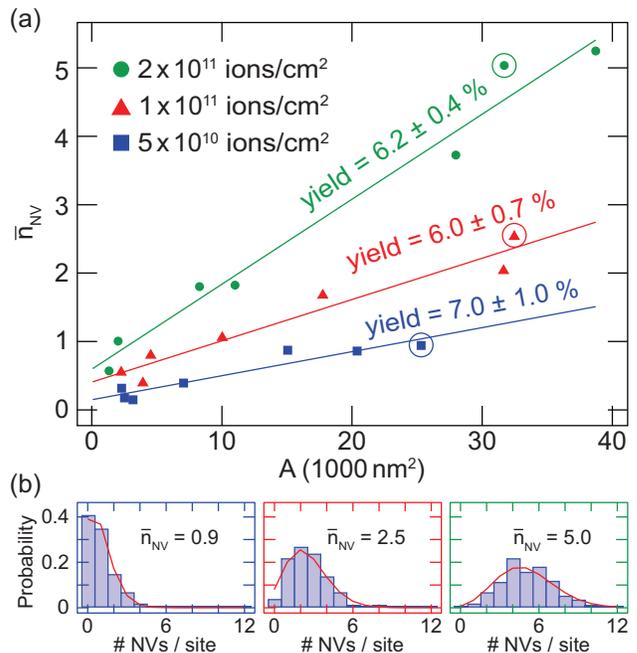}
\caption{ (Color online) (a) Average number of NV centers, $\bar{n}_\mathrm{NV}$, as a function of effective implantation aperture area, $A$, for three exposures of $^{15}$N$^+$, each with a different ion fluence. All three exposures result in a yield of $6-7\%$. (b) Population distributions of NV centers obtained from the three different ion fluences and aperture diameters that are indicated by the circles in (a). Solid lines are Poisson distributions for a given $\bar{n}_\mathrm{NV}$.}
\end{center}	
\end{figure}

From linear fits to the results, we obtain ion-to-NV conversion yields of $6.2 \pm 0.4\%$, $6.0 \pm 0.7\%$ and $7.0 \pm 1.0\%$ for average ion fluences of $2\times 10^{11}$, $1\times 10^{11}$, and $5\times 10^{10} \unit{ions/cm^2}$, respectively. We also found similar yields on a second sample that was implanted with an average ion fluence of $2\times 10^{11} \unit{ions/cm^2}$ (data not shown). Within each array of the same ion exposure and aperture diameter, the NV population distribution follows Poisson statistics, as shown in Fig.\ 3(b). These results demonstrate that we can reliably tune the average number of NV centers in each aperture.

\begin{figure} [tr!]
\begin{center}
		\includegraphics[width=\columnwidth]{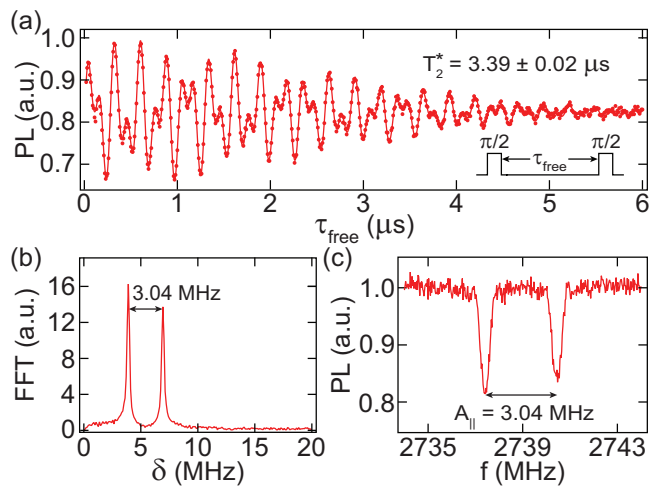}
\caption{ (Color online) (a) Ramsey fringes: PL plotted as a function of free precession time $\tau_\mathrm{free}$. A fit to the data (solid line) yields a dephasing time $T_2^*=3.39 \pm 0.02\mathrm{\ \mu s}$. (b) Fourier transform of the Ramsey data indicating two modulation frequencies separated by the hyperfine splitting $A_\parallel = 3.042 \pm 0.004 \mathrm{\ MHz}$. (c) Pulsed ODMR: PL plotted as a function of MW frequency $f$, showing $A_\parallel = 3.04 \pm 0.01 \mathrm{\ MHz}$.}
\end{center}	
\end{figure}

To probe the coherence of the implanted NV centers, we focus on implantation sites containing single centers and perform optically detected magnetic resonance (ODMR) by driving microwave (MW) signals through a $25\unit{\mu m}$ wire placed across the diamond surface. Working at a moderate magnetic field of $B \sim 50 \unit{G}$ aligned along the [111] direction, we can distinguish the NV centers with [111] orientation from those with other orientations using continuous wave ODMR. We then evaluate the quality of these NV centers by measuring the dephasing time, $T_2^*$, using a Ramsey experiment.\cite{Childress_Science_2006}

The Ramsey experiment is performed by first defining a two-level system using the $m_s=0$ and $m_s=-1$ states. After preparation of the electronic spin in $m_s=0$ by optical pumping, we apply a $\pi/2$-pulse to prepare a superposition state and then allow the state to evolve freely for a time interval $\tau_\mathrm{free}$. We then apply a second $\pi/2$-pulse to convert the coherence into population, followed by optical readout. Figure 4(a) shows the measured PL as a function of $\tau_\mathrm{free}$. The data display fast oscillations that decay with a characteristic timescale $T_2^*=3.39 \pm 0.02\mathrm{\ \mu s}$, a value that is comparable to naturally occurring NV centers in high quality, non-isotopically purified diamond.\cite{Childress_Science_2006,vanderSar_Nature_2012}

The fast oscillations in Fig.\ 4(a) are the result of the MW pulse being detuned by $\sim 5.5$ MHz from the $m_s=0$ and $m_s=-1$ transition. In our case, there are two hyperfine transitions associated with the $I=\frac{1}{2}$, $^{15}$N nuclear spin, leading to two different detunings and hence the beating of the signal in Fig.\ 4(a). By applying a Fourier transformation to the Ramsey data, we can extract the frequency of the two transitions relative to the MW drive. Figure 4(b) shows that the two hyperfine transitions are separated by a splitting of $A_\parallel = 3.042 \pm 0.004 \mathrm{\ MHz}$, consistent with previously reported values.\cite{Fuchs_PhysRevLett.101.117601,Felton_PhysRevB.77.081201} We also performed pulsed ODMR measurements by optically pumping the electronic spin to $m_s=0$ and applying a MW $\pi$-pulse with varying frequency, $f$, before optical readout.\cite{Jacques_PhysRevB.84.195204} When $f$ is on resonance with the transition from $m_s=0$ to $m_s=-1$, we see a reduction in PL from the NV center. With low MW powers (such that the transitions are minimally power broadened) we observe two hyperfine transitions yielding the same $A_\parallel = 3.04 \pm 0.01 \mathrm{\ MHz}$.

In conclusion, we have demonstrated a simple and reproducible way of forming NV centers via ion implantation. This work has critical applications for creating coupled NV systems and integrating them into larger scale QIP architectures. Recent studies of quantum gate operations involving multiple nuclear spins coupled to a NV center required the analysis of $\sim 3,300$ randomly-distributed NV centers to find the desired spin environment.\cite{Waldherr_Nature_2014} The controlled formation of NV centers at well-defined locations demonstrated here will allow for an efficient automation of such a characterization, while providing a desirable distribution of NV populations. In addition, since our positioning accuracy is limited only by electron beam lithography, our process allows for the placement of NV centers in nanopillars and on-chip optical resonators.\cite{Hausmann_NJP_2011, Hausmann_NanoLett_2012} This work advances the field toward realizing the full potential of the NV center for scalable QIP applications.

We thank Brandon Chance and Mike Souza for assistance with the perchloric acid etch. Supported by the Sloan and Packard Foundations, the National Science Foundation through the Princeton Center for Complex Materials (DMR-0819860) and CAREER award (DMR-0846341), and the Army Research Office (W911NF-08-1-0189).


\begin{thebibliography}{30}
\bibitem{footnote_amorphous}
While amorphous carbon pockets formed during implantations are not repairable with annealing, these pockets would only form at the damage threshold of $\sim$ 10 $\unit{vacancies/nm^3}$. This would require an implantation fluence of $> 1\times 10^{14} \unit{ions/cm^2}$ at 20 keV, which is more than two orders of magnitude above our maximum fluence.
\bibitem{footnote_SEM}
We note that we do not consider PL data from these sites after implantation due to the possibility of PMMA damage from SEM imaging.
\end{thebibliography}
\end{document}